# On the possible origin of the asteroid (1) Ceres

Yu. I. Rogozin[1,*]

**Abstract -** The last three decades the asteroid (1) Ceres is an object of the intensive ground-and space-based observations. A new unusual contributing to these studies represents the recent detection of localized sources of water vapour releasing from its surface at a rate about 6 kg s$^{-1}$ (Küppers et al 2014). A drastic distinction between asteroid (1) Ceres and nearest the large asteroid (4) Vesta in terms of their composition and appearance emphasizes an urgent state of a problem of the possible origin of Ceres in the main asteroid belt. By analogy with the early assumptions of some well-known astronomers of Mercury and Mars as the escaped satellites of their host planets we have put forward and semi-empirically have justified the hypothesis for the plausible origin of Ceres as the satellite of a disrupted planet in the past orbited the Sun of ~ 5 AU. The orbital location of this host of Ceres beyond the snow line of the Solar System explains the formation of the icy mantle of Ceres, which appears may be a water vapour source.

## 1. Introduction

Asteroid (1) Ceres - or dwarf planet Ceres depending on used definition - represents a peculiar object in the main asteroid belt (MAB) with a mass ~ 1/3 of the total mass of MAB among millions other asteroids and simultaneously a unique for asteroids of MAB hydrostatic equilibrium shape and homogeneous intact surface. Thus it is a subject of many studies from accurate data on its sizes and density (Thomas et al 2005) to suggesting observations to be made the *Dawn* mission (Russell & Raymond 2011). In so doing the issue of the origin of Ceres appears to be the clue to understanding the formation of the Solar System and it to date remains open. As noted previously, "establishing Ceres' birthplace will be necessary to fully understand its context" (McCord et al 2011).

In this connection the recent report of Küppers et al (2014) on the first observations releasing water vapour from the surface of asteroid (1) Ceres at a rate at least $10^{26}$ molecules, or 6 kg, being produced per second, originating from localized sources, has attracted considerable attention. Results of described observations naturally cause a statement of some questions, which clearly were formulated in comments to this article by H. Campins and C. M. Comfort (2014). These questions reduces to following: 1) Why asteroids (1) Ceres and (4) Vesta are so different on their composition and appearance (density of Ceres is 2.077 g cm$^{-3}$ but that of Vesta is 3.456 g cm$^{-3}$ and in addition the entire surface of Vesta covered with volcanic eruptions whereas Ceres has uniform icy surface)?; 2) Why did Ceres form with (and why does it still contain) more water than Vesta? It is most likely that Ceres formed in a colder outer region of the nascent Solar System than Vesta, beyond the snow line - the distance from the young Sun at which temperatures were low enough for water to form ice. But this hypothesis raises the question of why Ceres and Vesta are so close to each other now respectively approximately 2.8 and 2.4 AU.

Here, we will attempt to justify foregoing conjecture for a formation of Ceres in a colder remote the Sun region and to support a possible reason for its transition to MAB. To solve this problem we now use a double approach to determining the former location of Ceres in the early Solar System.

___________________________
[1]VEDA LLC, PO Box 85, 125368, Moscow, Russia
*e-mail: yrogozin@gmail.com
Submitted for publication

## 2. Analysis and results

First, we will refer to concept of mean motion resonance as applied to the planets of the inner Solar System. It is known orbital periods of planets Mercury and Venus respectively 0.2408 and 0.6152 years are locked near resonance 5:2. In so doing following the long remembered hypothesis (Van Flandern & Harrington 1976) we admit Mercury as the escaped former satellite of Venus. Moreover, we admit Mars with its orbital period 1.8808 year also as escaped satellite that is locked in the same resonance 5:2 to the perish planet Phaeton being orbited in the past in compliance with the Titius-Bode rule at 2.8 AU of the Sun with orbital period 4.6853 year. We consider this assumption justified as by recent simulations of Raymond et al (2009) in asteroid belt may have been stranded massive planet embryos. As objective evidence for a conceptual possibility to treat Mars as escaped in the past satellite of perish planet Phaeton orbited the Sun at 2.8 AU there can consider recent published conclusion that "most of Mars' building blocks consists of material that formed in the 2 AU to 3 AU region"(Brasser 2013).

In regard to Ceres we would to recall the old work (Kelly & Gaffey 1996) where the probability for the origin of Ceres in MAB as an interloper has been noted. Taking into account similar orbital features (large eccentricities and inclinations) to Mercury and Mars and strong distinction Ceres from other asteroids of MAB as well as its strange icy composition for the large celestial body inside of the snow line of the Solar System we may speculate a possibility for Ceres in the past be an escaped satellite of some disrupted planet being locked to it by analogy with foregoing pairs in the same resonance 5:2. In view of the present orbital period of Ceres 4.6 year that of its host may have been 11.5 year. Respectively, the semi-major axis of this disrupted planet might be 5.0949 AU.

With the intent to support this finding we now need to address to other approach. As known, the dwarf planet Ceres is similar to the trans-Neptunian dwarf planets (TNDPs) on own orbital parameters (large eccentricities and inclinations) and thus it may be proposed an escaped satellite similarly these TNDPs treated as former satellites of Neptune (Rogozin 2012) that could be the newcomer in the Solar System (Sumi et al 2011). Based on this, Ceres appears may have some common with them the regularities. Specifically, we find the empirical rule for inclinations of TNDPs $i$ in such form:

$$i = \frac{\theta}{2\pi} \frac{R_n}{R_N} \text{ (rad)} \qquad (1)$$

where $R_n$ and $R_N$ are semi-major axes respectively of $n$-th TNDP and Neptune, $\pi$ (=3.1416) and $\theta$ (=2.1390) are the mathematical constants with introduced here new constant $\theta$, which fit before unknown the mathematical identity $\pi \Phi \theta = 4 e$. Here, $\Phi$ (=1.6180) and $e$ (=2.7183) are known mathematical constants. In view of 1 rad = $57.29578^0$ coefficient $\theta/2\pi = 19.5^0$. Starting from foregoing our assumptions we admit the application of equation (1) to find the presently unknown orbital characteristics of Ceres' host (that we named Yurus). As an illustration, we used TNDPs Eris, Haumea, and Makemake, for which from this equation follows computed values $i$ equal respectively $43.88^0$, $28.10^0$, and $29.69^0$ versus observed values respectively $44.19^0$, $28.22^0$, and $28.96^0$. Now, using the present orbital data of Ceres $i = 10.586^0$ and $R_n = 2.765979$ AU (according to www. johnstonsarchive.net) from (1) follows the semi-major axis of Yurus should be 5.0951 AU, i.e. practically coincide with foregoing value $R_n$. This identity of the results derived from two different methods appears confirm our hypothesis on the origin Ceres in colder region of the Solar System. It is believed that the disruption of planet Yurus and the forced transition Ceres to MAB can result from a great gravitational effect of Jupiter at its inward migration to the Sun in the past.

**3.** Conclusions

The orbital location of this host of Ceres beyond the snow line of the Solar System explains a formation the icy mantle of Ceres, which appears may be a water source. The occurrence water vapour from the surface of Ceres could be associated with melting of some sites of icy mantle of Ceres at the expense of increase of the tidal influence of the Sun on its interior more than 3 times versus its former location remote from the Sun at simultaneous surface warming on its new orbit that became much closer to the Sun.

Although an accurate determining the physical properties of planet Yurus is beyond the scope of this Letter we yet point out that its mass appears could be ~ $10^{25}$ kg and its average density could be equal to that of sea ice. Past existence of such large disrupted icy planet appears give an insight into intermittent flows of briny water, recently observed on Mars (McEwen et al 2014), the temporary existence of the deep oceans and lakes on it (Masse et al 2014), and the origin of the vast masses of sea water on the Earth.